\documentclass[twocolumn]{aa}
\usepackage{graphicx}
\usepackage{natbib}
\bibpunct{(}{)}{;}{a}{}{,}
\usepackage{txfonts}
\begin{document}
   \title{X-ray continuum properties of GRB afterglows observed by XMM-Newton and Chandra}

   \author{B. Gendre \inst{1}, A. Corsi \inst{1}$^,$~ \inst{2}$^,$~ \inst{3}, \and L. Piro \inst{1}
          }

   \offprints{B. Gendre}

   \institute{IASF/INAF Rome, via fosso del cavaliere 100, 00133, Roma, Italy\\
              \email{bruce.gendre@rm.iasf.cnr.it, alessandra.corsi@rm.iasf.cnr.it, luigi.piro@rm.iasf.cnr.it}
     \and
         INFN - Sezione di Roma c/o Dipartimento di Fisica $-$ Universit\'a degli Studi di Roma "La Sapienza"
         Piazzale A. Moro 5, 00185 Roma, Italy
    \and
          Universit\'a degli Studi di Roma "La Sapienza", Piazzale A. Moro 5, 00185, Roma, Italy
             }

   \date{Received --; accepted --}

   \abstract{

We present a catalog of XMM-Newton and Chandra observations of gamma-ray burst (GRB) afterglows, reduced in a common way using the most up-to-date calibration files and software. We focus on the continuum properties of the afterglows. We derive the spectral and temporal decay indices for 16 bursts. We place constraints on the burst environment and geometry. A comparison of the fast XMM-Newton follow-up and the late Chandra observations shows a significant difference in those parameters, likely produced by a transition from jet expansion taking place between two and ten days after the burst. We do not observe a significant shrinking of the luminosity distribution when we correct for beaming;  more burst observations are needed to confirm this result.
We also compare our results with those obtained by BeppoSAX and SWIFT; there is no strong discrepancy between the afterglow fluxes observed with these satellites when we carefully take into account the different median observation time of each observatory.

   }

\titlerunning{XMM-Newton and Chandra GRB X-ray afterglows}
\authorrunning{Gendre et al.}

   \maketitle
%

\section{Introduction}

Discovered at the end of the 60s \citep{kle73}, Gamma-Ray Bursts (GRBs) remained mysterious objects for decades. The first precise localization of long GRB \citep{dez96,kou93} provided by BeppoSAX \citep{piro96} led to the detection of X-ray \citep{cos97} and optical afterglows \citep{van97}, and finally to the measurement of the distance of these objects \citep{met97}. The afterglow emission is well described by the fireball model \citep{ree92, mes97, pan98} and provides powerful diagnostics of the close environment of the burst. In the framework of the fireball model, the afterglow spectral and temporal evolution can be used to constrain the density profile of the environment. Many observations point out a possible association between GRBs and supernovae \citep[e.g.][]{sta03, hjo03}, indicating that the long GRB progenitor may be a massive star. In such a case the surrounding medium is then expected to be the stellar wind arising from the star \citep{che99}. However most of the measurements are compatible with a constant density profile and in some cases exclude a wind profile \citep[e.g;][]{pan02}. Only in a few cases does the contrary hold \citep[e.g.][]{gen04, piro05}.

A large number of datasets available for a systematic study in X-ray is available. Some observations were published in single papers, other announced in GCN, and some not published at all. A uniform sample of afterglows, obtained by using the latest calibration files, is particularly required when one wants to use the general properties of the afterglows to constrain models. We have initiated a re-analysis of all X-ray afterglows observed so far, focusing on the burst environment properties. Following the release of the BeppoSAX results \citep{dep05}, we present in this article the XMM-Newton and Chandra afterglow observations carried out as of October 2004. We will focus on the continuum properties, and we will not discuss any line detections within the spectra. In a future paper (Gendre et al., in preparation), we will expose a methodology to detect lines, to assess their significance and we will discuss the features that we have detected. The light curves and spectra will be made available through a local web page\footnote{http://grb.rm.iasf.cnr.it/catalog}.

This article is organized as follows. In Sec. \ref{sec_ana} we present the data reduction. In Sec. \ref{sec_obs} we list our results. In Sec. \ref{sec_discu} we discuss these results and compare them with the BeppoSAX catalog, before providing our conclusions.


\section{Data reduction and analysis}
\label{sec_ana}

We retrieved all public data available for GRBs that occurred before the 1st of October 2004, from the archives of XMM-Newton\footnote{http://xmm.vilspa.esa.es/external/xmm\_data\_acc/xsa/index.shtml} and Chandra\footnote{http://cda.harvard.edu/chaser/mainEntry.do} observations. The complete list of retrieved observations is indicated in Table \ref{table1}. These observations were calibrated using the most up-to-date software available as of October 2004. 
 For the XMM-Newton data, we used the SAS version 6.0. We focused on the data from the EPIC (MOS and PN) instruments \citep{str01, tur01}. The calibrations were done using the tasks {\it emchain} and {\it epchain}. For the Chandra data, we used CIAO version 3.1 and the calibration database CALDB version 2.28. The data were processed using the tasks {\it acis\_process\_events} and the specific tasks used to obtain calibrated grating data\footnote{See http://cxc.harvard.edu/ciao3.1/threads/spectra\_hetgacis/ and http://cxc.harvard.edu/ciao3.1/threads/spectra\_letgacis/}. The events were filtered using all provided GTI and standard filtering criteria (GRADE==[0, 2, 3, 4, 6] for Chandra data, (FLAG==0\&\&PATTERN$<=$4) for XMM-Newton EPIC-PN data, (\#XMMEA\_EM\&\&PATTERN$<=$12) for XMM-Newton EPIC-MOS). The filtered event files were checked for flaring background activities. We removed any period of such activity using a very strict condition (a flare is defined as an increase by a factor of at least 5 of the mean background count rate during the observation, and we removed the detected flare including the 250 seconds that precede and follow this period). When less than 3000 seconds of observation remain within the cleaned event file, we discarded the observation. The remaining events were used to extract spectra and light curves.

For non-grating data and the zero order grating data, we extracted spectra and light curves using circle regions. The radius of these regions were chosen to contain at least 90\% of the counts and take into account any possible neighboring source. The spectral and temporal backgrounds were extracted using a larger circular area free of sources at the same off-axis angle (except in the case of the zero order grating data, where we extracted this spectrum using an annulus centered on the source with inner and outer radii of about 10 and 25 pixels respectively). The XMM-Newton spectral background presents spatial variations \citep[see][]{ehl04}. We have taken this into account by choosing a position that presents a spectral background similar to the source region, in order to avoid spurious emission or absorption features. For the other orders of grating data, we used the regions created by the grating processing tasks\footnote{see the CIAO threads web pages}. When grating data were available (either LETG or HETG), we obtained the spectral results by simultaneously fitting the zero order spectrum and the $+1$ and $-1$ orders added together. In such a case, we have taken into account the cross-calibration uncertainties, which are about 20\% (see the CIAO calibration web pages), by adding to the spectral model a multiplicative constant left free to vary within the range 0.8-1.2. The spectral analysis was done using XSPEC version 11.3.1 \citep{arn96}. For the temporal analysis we used FTOOLS version 5.3 and a customized IDL script \citep[see][]{dep05}. We binned all data to contain at least 20 source photons within each bin in order to use the $\chi^2$ statistic. For the spectral study, we used a model of an absorbed power law (with a galactic absorption and a host redshifted absorption). When the redshift of the host was unknown, we used the canonical value of z=1. For those afterglows with too few counts detected to produce a meaningful spectral fitting, we used only a galactic absorbed power law with a spectral energy index of 1 to calibrate the count to flux conversion factor. For the Chandra grating data, we used only the order 0 of the spectra to extract the light curve, as this was found to optimize the signal-to-noise ratio and thus reduce the uncertainties.

\begin{table*}
\caption{List of observations retrieved from the archives with some basic information (configuration of the instruments, instrument that detected the prompt emission). We also report the first refereed publication on the X-ray afterglow.}
\label{table1}
\centering
\begin{tabular}{cccccccc}
\hline\hline
GRB                 & Instrument           & archive    & Prompt   & Fluence (band)           & Afterglow   & Redshift & Ref\\
                    &                      & Obs-ID     & detection& erg.cm$^{-2}$ (keV)      & Detection   &          &    \\
\hline
\object{GRB 991216} & Chandra ACIS-S, HETG & 596        & BATSE    & $2\times10^{-4}$(20-300) & X, opt, rad &  1.02    & 1  \\
\object{GRB 000210} & Chandra ACIS-S       & 602        & BeppoSAX &$6.1\times10^{-6}$(40-700)& X, rad      &  0.846   & ---\\
\object{GRB 000926} & Chandra ACIS-S       & 598, 1047  & Ulysses  &$2.2\times10^{-5}$(25-100)& X, opt, rad &  2.066   & 2  \\
\object{GRB 010222} & Chandra ACIS-S       & 2424, 1048 & BeppoSAX &$9.25\times10^{-5}$(40-700)&X, opt, rad &  1.477   & ---\\
\object{GRB 011030} & Chandra ACIS-S       & 3411, 3412 & BeppoSAX &$1.13\times10^{-6}$(40-700)&rad         &  $<$3    & ---\\
\object{GRB 011130} & Chandra ACIS-I       & 2823       & HETE-2   & ---                      & ---         &  ---     & ---\\
\object{GRB 020127} & Chandra ACIS-I       & 3436, 3437 & HETE-2   & ---                      & X, rad      &  ---     & ---\\
\object{GRB 020321} & Chandra ACIS-S       & 3477       & BeppoSAX &$3.0\times10^{-6}$(40-700)& ---         &  ---     & ---\\
\object{GRB 020405} & Chandra ACIS-S, LETG & 2825       & Ulysses  & $3\times10^{-5}$(25-100) & X, opt, rad & 0.69     & 15 \\
\object{GRB 020427} & Chandra ACIS-S       & 3493, 3494 & BeppoSAX &$<2.9\times10^{-7}$(40-700)&X           & ---      & 3  \\
\object{GRB 020531} & Chandra ACIS-I       & 2824, 3673 & HETE-2   & $1\times10^{-7}$(8-40)   & ---         & ---      & ---\\
\object{GRB 020813} & Chandra ACIS-S, HETG & 4364       & HETE-2   &$3.8\times10^{-5}$(25-100)& X, opt, rad & 1.25     & 4  \\
\object{GRB 021004} & Chandra ACIS-S, HETG & 4381, 4409 & HETE-2   &$3.2\times10^{-6}$(7-400) & X, opt, rad & 2.3      & 4  \\
\object{GRB 021201} & Chandra ACIS-I       & 3784, 3785 & IPN      &$2\times10^{-7}$(25-100)  & ---         & ---      & ---\\
\object{GRB 030226} & Chandra ACIS-S       & 4425       & HETE-2   &$5.7\times10^{-6}$(30-400)& X, opt      &  1.98    & 5  \\
\object{GRB 030328} & Chandra ACIS-S, LETG & 4432       & HETE-2   &$3.0\times10^{-5}$(30-400)& X, opt      & 1.52     & ---\\
\object{GRB 030528} & Chandra ACIS-S       & 3920, 3921 & HETE-2   &$4.8\times10^{-6}$(30-400)& X, opt      & ---      & 6  \\
\object{GRB 030723} & Chandra ACIS-S -I    & 3918, 3922 & HETE-2   &$0.2\times10^{-7}$(30-400)& X, opt      & ---      & 18 \\
\object{GRB 031220} & Chandra ACIS-I       & 3905, 3906 & HETE-2   &$1.9\times10^{-6}$(25-100)& X, opt      & ---      & 7  \\
\object{GRB 040701} & Chandra ACIS-I       & 4661, 4662 & HETE-2   &$4.5\times10^{-7}$(2-25)  & X           & 0.2146   & ---\\
\object{GRB 040812} & Chandra ACIS-S       & 5364, 5365 & INTEGRAL &$2.5\times10^{-7}$(20-200)& X           & ---      & ---\\
\hline
\object{GRB 001025A}& XMM-Newton           &0128530301  & Ulysses  &$3.2\times10^{-6}$(25-100)& ---         &   ---    & 8  \\
\object{GRB 010220} & XMM-Newton           &0111520301  & BeppoSAX &      ---                 & ---         &   ---    & 8  \\
                    &                      &0111520401  &          &                          &             &          &    \\
\object{GRB 011211} & XMM-Newton           &0094380101  & BeppoSAX &$3.7\times10^{-6}$(40-700)& X, opt      &  2.14    & 9  \\
\object{GRB 020321} & XMM-Newton           &0008820401  & BeppoSAX &$3.0\times10^{-6}$(40-700)& ---         &  ---     & ---\\
\object{GRB 020322} & XMM-Newton           &0110980301  & BeppoSAX &      ---                 & X, opt      &  ---     & 10 \\
                    &                      &0128530801  &          &                          &             &          &    \\
\object{GRB 030227} & XMM-Newton           &0128531201  & INTEGRAL &$8.8\times10^{-7}$(20-200)& X, opt      &  ---     & 11 \\
\object{GRB 030329} & XMM-Newton           &0128531401  & HETE-2   &$1\times10^{-4}$(30-400)  & X, opt, rad &  0.168   & 12 \\
                    &                      &0128531501  &          &                          &             &          &    \\
                    &                      &0128531601  &          &                          &             &          &    \\
\object{GRB 031203} & XMM-Newton           &0158160201  & INTEGRAL &$1.5\times10^{-6}$(20-200)& X, opt, rad &  0.105   & 13 \\
                    &                      &0163360201  &          &                          &             &          &    \\
\object{GRB 040106} & XMM-Newton           &0158160401  & INTEGRAL &$8.3\times10^{-7}$(20-200)& X, opt      &  ---     & 14 \\
\object{GRB 040223} & XMM-Newton           &0158160601  & INTEGRAL &$4.4\times10^{-7}$(20-200)& X           &  ---     & 17 \\
\object{GRB 040827} & XMM-Newton           &0164570401  & INTEGRAL & ---                      & X, opt      &  ---     & 16\\
\hline
\end{tabular}

N.B. References are : 1 : \citet{piro00}, 2 :  \citet{piro01}, 3 : \citet{ama04}, 4 : \citet{but03}, 5 : \citet{klo04}, 6 : \citet{rau04}, 7 : \citet{mel05}, 8 : \citet{wat02a}, 9 : \citet{ree02}, 10 : \citet{wat02b}, 11 : \citet{mer03}, 12 : \citet{tie03}, 13 : \citet{wat04}, 14 : \citet{gen04}, 15 : \citet{mir03}, 16 : \citet{del05}, 17 : \citet{gly05}, 18 : \citep{but05a}.

\end{table*}

\section{Observation results}
\label{sec_obs}

We list in Table \ref{table3} our results. We indicate undetected afterglows (we define as afterglow a fading bright source, or a single source within the error box too faint to constrain its decay) with a 'U' flag. For these afterglows, we indicate a flux upper limit assuming the source was on-axis (when there is no afterglow detected at other wavelengths) or at the position of the detected optical/radio afterglow. Bursts detected but with not enough counts to have a good constraint on the spectral index are indicated with an 'F' flag. We have derived a flux for those afterglows using the standard model indicated in Sec. \ref{sec_ana}. For the remaining bursts, we indicate if the observation was done by XMM-Newton ('X' flag) or by Chandra in grating mode ('G' flag) or imaging mode ('I' flag). In the remainder of this paper, all errors are quoted at the 90\% confidence level (except when another confidence level is indicated). Fluxes are given unabsorbed and within the 2.0-10.0 keV X-ray band. Readers interested in using our data are invited to retrieve them through the dedicated web page$^1$.

\begin{figure}
\centering
\includegraphics[width=9cm]{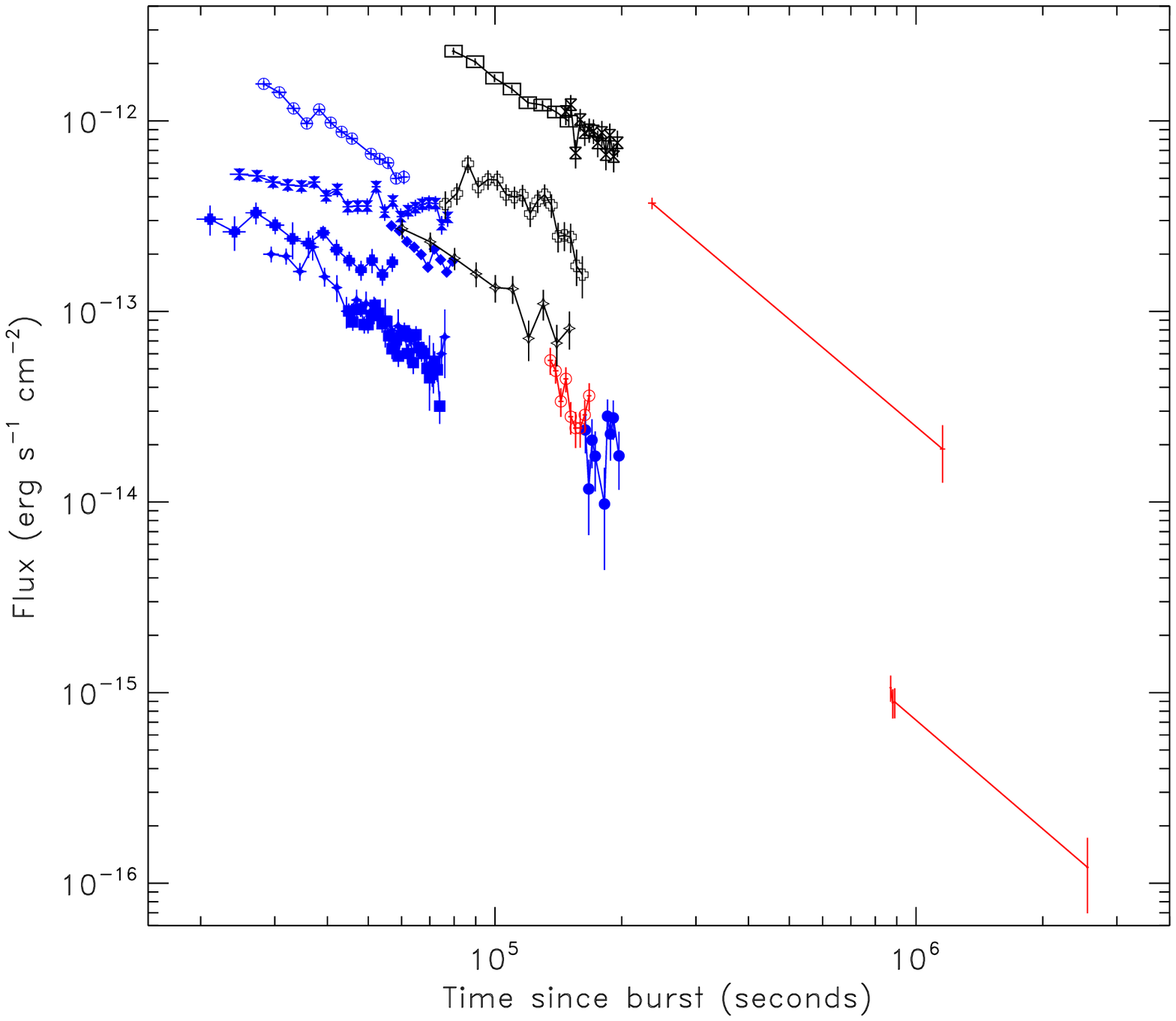}
\caption{\label{figure_lc_tot}Light curves of the bursts with a good constraint on both the spectral and temporal indexes. We have indicated in blue the XMM-Newton bursts, in black the Chandra Grating bursts and in red the Chandra Imaging bursts (see electronic version for color).}
\end{figure}

\begin{table*}
\caption{GRB X-ray afterglow results. For each afterglow, we give the source name, the exposure duration and the net observation duration, together with the results of the spectral and temporal analysis. We quote the absorption (measured at the host redshift or at a value of z=1 when the host redshift is unknown) in excess to the galactic absorption. The X-ray absorption excess upper limits are given at the 90 \% confidence level. We also indicate the number of detected counts for the source as a proxy of the detection confidence level.}
\label{table_chandra_grating}
\label{table_chandra_imaging}
\label{table_xmm}
\label{table3}
\label{table4}
\centering
{\scriptsize
\begin{tabular}{ccccccccccc}
\hline\hline

Source  & Flag & Time             & Exposure & Net      & Detected & Temporal & Energy   & Spectral & Excess of & Flux\\
name    &      &since      burst  & duration & duration & count   & decay    & spectral & fit & absorption & \\
        &      &(days)            &  (ksec)  & (ksec)   &number& &index&$\chi_\nu^2$ (d.o.f.)&($10^{21}$ cm$^{-2}$)&erg s$^{-1}$ cm$^{-2}$\\
\hline
\object{GRB 991216}&G&1.52 & 9.8  & 9.8  &1305&  (0.9)        &  0.7 $\pm$ 0.2 & 0.94 (56) & $< 9.6$    &1.92 $\pm$ 0.08 $\times 10^{-12}$\\
\object{GRB 000210}&I&0.84 & 10.0 & 10.0 &533 & 1.38$\pm$0.03 & 0.9  $\pm$ 0.2 & 1.0 (15)  & 4.7 $\pm$ 1.3& $2.7 \pm 0.2 \times 10^{-13}$\\
\object{GRB 000926}&I&2.67 & 10.0 & 10.0 &307 &1.9  $\pm$ 0.4& 0.7  $\pm$ 0.2 & 1.5 (8)   & $< 3.4$      &$1.06 \pm 0.07 \times 10^{-13}$\\
\object{GRB 001025A}&X&1.88& 40.8 & 33.4 &911 & (1.2)        & 1.8  $\pm$ 0.4 & 0.90 (13) & 6 $\pm$ 3     & $2.0 \pm 0.2 \times 10^{-14}$\\
\object{GRB 011030}&I&10.00& 50.0 & 29.6 &248 & 1.9  $\pm$ 0.4& 0.7  $^{+2.1}_{-1.0}$ & 0.1 (6)& $< 39$   & $ 5 \pm 1.2 \times 10^{-14}$\\
\object{GRB 011130}&U& 9.77& 30.2 & 30.2 & ---&       ---    &     ---        &  ---      &    ---        & $<0.6\times 10^{-15}$    \\
\object{GRB 011211}&X& 0.50& 33.6 & 33.6 &3589& 2.1 $\pm$ 0.3& 1.3  $\pm$ 0.1 & 1.20 (67) &    $<$4.1     & $7.3 \pm 0.2 \times 10^{-14}$\\
\object{GRB 020127}&F& 4.05& 10.0 & 10.0 &26  &       ---    &     ---        &  ---      &    ---        & 2.1 $\pm 0.5 \times 10^{-14}$\\
                   & &14.53& 10.1 & 10.1 &13  &       ---    &     ---        &  ---      &    ---        & 0.6 $\pm 0.2 \times 10^{-14}$\\
\object{GRB 020321}&F?&0.43& 50.0 & 32.0 &227 &       ---    &     ---        &  ---      &    ---        & see text                 \\
\object{GRB 020322}&X& 0.62& 29.0 & 28.4 &4226& 1.3 $\pm$ 0.3& 1.1  $\pm$ 0.1 & 1.00 (178)& 6.5 $\pm$ 0.8 &$2.11 \pm 0.05 \times 10^{-13}$\\
\object{GRB 020405}&G&1.68 & 51.2 & 51.2 &2086& 1.6 $\pm$ 0.9 &  1.0 $\pm$ 0.2 & 1.24 (75) &5 $\pm$ 2     &8.9 $\pm$ 0.3 $\times 10^{-13}$\\
\object{GRB 020427}&F& 9.06& 13.9 & 13.9 &55  &       ---    &     ---        &  ---      &    ---        & 1.0 $\pm 0.2 \times 10^{-14}$\\
                   & &17.00& 14.8 & 14.8 &24  &       ---    &     ---        &  ---      &    ---        & 0.4 $\pm 0.1 \times 10^{-14}$\\
\object{GRB 020531}&U& 5.14& 20.1 & 20.1 & ---&       ---    &     ---        &  ---      &    ---        & $<1.2\times 10^{-15}$    \\
\object{GRB 020813}&G&0.86 & 78.1 & 78.1 &8041& 1.4 $\pm$ 0.2 & 0.83 $\pm$ 0.06& 0.93 (364)& $< 2.9$    &1.63 $\pm$ 0.03 $\times 10^{-12}$\\
\object{GRB 021004}&G&0.85 & 88.1 & 88.1 &2656& 1.2 $\pm$ 0.3 & 1.01 $\pm$ 0.07& 1.08 (112)& $< 2.8$      &4.0 $\pm$ 0.2 $\times 10^{-13}$\\
\object{GRB 021201}&U& 8.73& 20.0 & 20.0 & ---&       ---    &     ---        &  ---      &    ---        & $<1.2\times 10^{-15}$    \\
\object{GRB 030226}&I&1.54 & 40.0 & 40.0 &371 &2.7  $\pm$ 1.6&0.9 $\pm$ 0.3 & 1.1 (10)&6.6$^{+4.1}_{-3.3}$& $ 3.6 \pm 0.2 \times 10^{-14}$\\
\object{GRB 030328}&G&0.64 & 98.1 & 98.1 &1575& 1.6 $\pm$ 0.3 & 1.1  $\pm$ 0.2 & 0.79 (39) & $< 4.0$    &1.70 $\pm$ 0.08 $\times 10^{-13}$\\
\object{GRB 030329}&X&37.04& 244.9& 143.5&639 &    (2)       & 1.0  $\pm$ 0.2 & 0.98 (22) &    $<$0.5     & $1.8 \pm 0.2 \times 10^{-14}$\\
\object{GRB 030528}&F& 5.97& 26.1 & 26.1 &44  &       ---    &     ---        &  ---      &    ---        & 7.8 $\pm 1.3 \times 10^{-15}$\\
                   & &11.80& 20.3 & 20.3 &--- &       ---    &     ---        &  ---      &    ---        & $< 3.2 \times 10^{-15}$      \\
\object{GRB 030723}&F& 2.14& 24.8 & 24.8 &76  &       ---    &     ---        &  ---      &    ---        & $1.4 \pm 0.7 \times 10^{-14}$\\
                   & &12.69& 84.1 & 84.1 &70  &       ---    &     ---        &  ---      &    ---        & $2.3 \pm 1.7 \times 10^{e-15}$\\
\object{GRB 031203}&X& 0.26& 112.8& 92.2 &6261& 0.5 $\pm$ 0.1& 0.8  $\pm$ 0.1 & 1.04 (244)& 2.6 $\pm$ 0.7 &$3.95 \pm 0.08 \times 10^{-13}$\\
\object{GRB 031220}&F& 5.62& 40.2 & 40.2 &31  &       ---    &     ---        &  ---      &    ---        & 5.2 $\pm 1.3 \times 10^{-15}$\\
                   & &28.47& 19.7 & 19.7 &--- &       ---    &     ---        &  ---      &    ---        & $< 1.5 \times 10^{-15}$       \\
\object{GRB 040106}&X&0.23 & 44.6 & 37.0 &15738&1.4 $\pm$ 0.1& 0.49 $\pm$ 0.04& 1.07 (515)&    $<$0.5     & $9.1 \pm 0.2 \times 10^{-13}$\\
\object{GRB 040223}&X& 0.20& 45.2 & 14.3 &1230&0.7 $\pm$ 0.3& 1.6 $\pm$ 0.3 &1.04 (43)&(45-88) $\pm$ 16$^a$& $2.3 \pm 0.2 \times 10^{-13}$\\
\object{GRB 040701}&F& 7.77& 20.0 & 20.0 &68  &       ---    &     ---        &  ---      &    ---        & 1.2 $\pm 0.2 \times 10^{-14}$ \\
                   & &16.50& 19.0 & 17.5 &39  &       ---    &     ---        &  ---      &    ---        & 0.8 $\pm 0.2 \times 10^{-14}$ \\
\object{GRB 040812}&F& 5.04& 10.2 & 10.2 &58  &       ---    &     ---        &  ---      &    ---        & 1.6 $\pm 0.3 \times 10^{-14}$ \\
                   & &10.16& 10.2 & 10.2 &42  &       ---    &     ---        &  ---      &    ---        & 1.2 $\pm 0.2 \times 10^{-14}$ \\
\object{GRB 040827}&X&0.27 & 52.3 & 39.3 &2899& 1.4 $\pm$ 0.3& 1.3  $\pm$ 0.1 & 1.07 (116)& 8.6 $\pm$ 1.5 &$1.25 \pm 0.04 \times 10^{-13}$\\
\hline
\end{tabular}}

Note : $^a$ : A high galactic density column affect this observation. See Sec. \ref{section_absorp} for details.

\end{table*}

There were 31 bursts followed by XMM-Newton or Chandra (\object{GRB 020321} was followed by XMM-Newton and Chandra). We obtained good constraints on the spectral and temporal indexes for 14 of these. For 3 other bursts (\object{GRB 991216}, \object{GRB 001025A}, \object{GRB 030329}), the decay index was not constrained but we obtained a good statistical error on the spectral index. For 8 bursts, the observed count rates did not allow us to derive good constraints on the spectral and decay indexes. For 3 bursts (\object{GRB 011130}, \object{GRB 020531}, \object{GRB 021201}), we did not detect any convincing afterglow. We discarded 3 burst (\object{GRB 010220}, \object{GRB 010222}, \object{GRB 030227}) observations. Except for the case of \object{GRB 011030}, all results reported here are consistent with previous publications (which are listed in Table \ref{table1}). In the special case of \object{GRB 011030}, \citet{har01} reported a very different flux, not consistent with the count rate observed for this source. An independent analysis made by \citet{ale05} gave results similar to ours. We show in Fig. \ref{figure_lc_tot} the light curves (in flux unit) of those bursts with a good constraint on both the spectral and decay indexes.

\subsection{Rejected observations}

\paragraph{\object{GRB 010220}}

A strong flaring background activity occurred during the XMM-Newton observation, which began 14.8 hours after the burst. Due to our strict flare rejection criterion, we discarded all the observation of this burst.

\paragraph{\object{GRB 010222}}

This observation was severely piled-up.
We decided not to perform data analysis and to conservatively discard this observation.

\paragraph{\object{GRB 030227}}

The observation files gave several warnings during the data processing by the SAS. After a check of the event files, we concluded that the calibration could not be correctly done for this observation. We conservatively discarded this observation.

\subsection{Particular afterglows}

\begin{table}
\caption{List of GRB afterglow fluxes extrapolated or interpolated 11 and 12 hours after the burst from their spectral and temporal parameters (assuming there is no evolution of these parameters). These two times are chosen to match the works by \citet{dep05} and \citet{ber05} about Beppo-SAX and SWIFT respectively. We have separated the Chandra Grating (top), Chandra Imaging (center) and XMM-Newton (bottom) samples due to their different observation times, and have removed \object{GRB 030329} due to its very late XMM-Newton observation.}             
\label{table_flux}      
\centering                          
\begin{tabular}{ccc}        
\hline\hline                 
GRB Name            & Flux                          &   Flux                    \\ 
                    &  11 hours                     &   12 hours              \\
                    & (erg cm$^{-2}$ s$^{-1}$)      & (erg cm$^{-2}$ s$^{-1}$) \\
\hline
\object{GRB 020405} & $7.1 \pm 0.3 \times 10^{-12}$ & $6.2 \pm 0.3 \times 10^{-12}$ \\
\object{GRB 020813} & $3.93\pm0.08 \times 10^{-12}$ & $3.48\pm0.07 \times 10^{-12}$ \\
\object{GRB 021004} & $8.4 \pm 0.5 \times 10^{-13}$ & $7.6 \pm 0.4 \times 10^{-13}$ \\
\object{GRB 030328} & $2.9 \pm 0.2 \times 10^{-13}$ & $2.5 \pm 0.2 \times 10^{-13}$ \\
\hline
\object{GRB 000210} & $6.7 \pm 0.3 \times 10^{-13}$ & $6.0 \pm 0.3 \times 10^{-13}$  \\
\object{GRB 000926} & $3.1 \pm 0.2 \times 10^{-12}$ & $2.7 \pm 0.2 \times 10^{-12}$  \\
\object{GRB 011030} & $1.8 \pm 0.5 \times 10^{-11}$ & $1.6 \pm 0.4 \times 10^{-11}$  \\
\object{GRB 030226} & $1.35\pm0.08 \times 10^{-12}$ & $1.10\pm0.07 \times 10^{-12}$ \\
\hline
\object{GRB 001025A}& $1.2 \pm 0.2 \times 10^{-13}$ & $1.1 \pm 0.2 \times 10^{-13}$  \\
\object{GRB 011211} & $1.72\pm0.05 \times 10^{-13}$ & $1.46\pm0.05 \times 10^{-13}$  \\
\object{GRB 020322} & $4.2 \pm 0.1 \times 10^{-13}$ & $3.8 \pm 0.1 \times 10^{-13}$  \\
\object{GRB 031203} & $4.5 \pm 0.1 \times 10^{-13}$ & $4.3 \pm 0.1 \times 10^{-13}$  \\
\object{GRB 040106} & $9.9 \pm 0.2 \times 10^{-13}$ & $9.0 \pm 0.2 \times 10^{-13}$  \\
\object{GRB 040223} & $2.3 \pm 0.2 \times 10^{-13}$ & $2.2 \pm 0.2 \times 10^{-13}$  \\
\object{GRB 040827} & $1.40\pm0.05 \times 10^{-13}$ & $1.26\pm0.05 \times 10^{-13}$  \\
\hline                           
\end{tabular}
\end{table}

\paragraph{\object{GRB 000210}}

We used the data from BeppoSAX to constrain the decay index. 

\paragraph{\object{GRB 001025A}}

 Two sources were detected within the burst error box, the brightest displaying a marginal decay ($\delta = 1.2 \pm 3.0$, which implies a decrease at the 66\% confidence level) while the dimmer does not feature any variation. \citet{wat02a} associated the bright source with the X-ray afterglow on the basis of this decay trend. We report on this bright source.

\paragraph{\object{GRB 011211}}

There was a change in the attitude of the satellite during the observation. As this change of attitude is not fully supported by the SAS and complicates the spectral analysis (due to the change of PN CCD-chip during the re-pointing), we discarded the first 5 kiloseconds of the observation (before the change of attitude). 

\paragraph{\object{GRB 020321}}

This burst was detected by BeppoSAX, who failed to detect its afterglow \citep{zan03}. It was followed-up by XMM-Newton and Chandra 10.3 and 240 hours after the burst respectively. We detected 82 sources within the field of view of XMM-Newton. Twelve of these sources are located in the Wide Field Camera (WFC) error box \citep{gan02}. The subsequent Chandra observation covered half of the field of view of XMM-Newton. One bright XMM-Newton source disappeared in the Chandra field of view, which was claimed to be the afterglow of \object{GRB 020321} by \citet{zan03}. This source has a flux of $1.4 \times 10^{-14} \pm 0.2 \times 10^{-14}$ erg cm$^{-2}$ s$^{-1}$ 17 hours after the burst, and an upper limit of $\sim 1.1 \times 10^{-15}$ erg s$^{-1}$ cm$^{-2}$ during the Chandra observation. On the other hand, only about 80 \% of the WFC error box is covered by Chandra, and in particular not all the bright XMM-Newton sources are in the Chandra field of view \citep[as also noted by][]{zan03}. Five XMM-Newton sources are located outside the field of view of Chandra, but inside the WFC error box. 
There were no other observations at other wavelenghts to confirm this likely afterglow, and we cannot exclude that this variable source may be not related to \object{GRB 020321}.

\paragraph{\object{GRB 030329}}

There were 3 late observations of this burst. The analysis was complicated by a neighboring0 source that contaminated the spectra of the afterglow. While this contamination is low and acceptable in the first and second observations, the last observation suffered strong contamination that prevented us from analyzing the data. Due to the late observation time, the decay index is not constrained using only the XMM-Newton data \citep[we refer the reader to][for the early RXTE observation]{tie03}.

We report in Table \ref{table_xmm} the spectral analysis of only the first observation. We did not detect any excess of absorption within the spectra of both observations. The spectral index is $\alpha = 0.85 \pm 0.21$ ($\chi^2_{\nu} = 0.71$, 10 d.o.f.) in the second observation, consistent with that measured in the first observation.

\paragraph{\object{GRB 031203}}

There were 2 observations of this burst, and we report in Table \ref{table_xmm} only the first observation results. The spectral index is $\alpha = 1.0 \pm 0.3$ ($\chi^2_{\nu} = 0.70$, 18 d.o.f.) for the second observation, consistent with that measured in the first observation. The decay index is compatible between the two observations. An expanding ring of X-ray, interpreted as the dust-scattered echo of the prompt emission was also detected during the first observation \citep{wat04}.

\paragraph{\object{GRB 031220}}

 \citet{mel05} have reported 2 possible candidates which have the same flux during the first TOO. We report only the proposed candidate, which was not detected during the second TOO. 

\paragraph{\object{GRB 040106}}

The light curve of the PN instrument displays a decay with an index of 1.4 $\pm$ 0.1. We note that \citet{mor05} reported a very high constraint on the decay index, using a summed light curve. We thus summed the light curves of the three EPIC instruments into a single light curve. We obtained a decay index of $\delta = 1.38 \pm 0.09$ (90\% confidence level). Using only the 1 sigma level \citep[as in][]{mor05}, the error drops to 0.04, i.e. similar to their value.

\paragraph{\object{GRB 040701}}

This burst was observed twice, but several sources within the position error box show variations. We report only the candidate proposed by \citet{fox04}.

\section{Discussion}
\label{sec_discu}

\subsection{Comparison between afterglows}

 Twenty four afterglows were securely detected (out of 28 observations), and one observation gave a doubtful afterglow detection. This is a detection efficiency of 85 \% (89 \% if the doubtful detection is indeed an afterglow). We discuss why some afterglows (\object{GRB 011130}, \object{GRB 020531}, and \object{GRB 021201}) were not detected in Sec. \ref{sec_geometry}, and propose as an explanation a jet effect. Our sample can be separated into the XMM-Newton, Chandra-grating and Chandra-imaging sub-samples. The selection criteria to perform an observation are different in each of these samples. In the Chandra-grating sample, most of the observations were triggered because of a bright prompt emission detected in X-ray instruments. These bursts were therefore selected to be bright (see Table \ref{table_flux}) in order to produce good quality high resolution spectra. The Chandra-imaging sample is mostly constituted of approved TOOs. Each of these TOOs have different goals (e.g. accurate localization, high redshift burst hunt), and one may consider this sample as not too biased. The XMM-Newton sample is constituted of bursts observed in the Director's Discretionary Time. These bursts (except \object{GRB 030329}) where rapidly followed because it was possible to observe them, without a-priori information on the flux level. One can consider these burst observations as randomly chosen within all the possible observations, and thus not biased. In the following, we have assumed that the Chandra-imaging and XMM-Newton samples do not present a selection bias, and we have excluded from the flux comparison the Chandra-grating sample, as these bursts were chosen to be very bright (and thus present a bias toward the bright end of the afterglow luminosity distribution). 

We list the fluxes, interpolated or extrapolated to a common time of 11 and 12 hours, of those bursts with a good constraint on the decay index in Table \ref{table_flux}. We list the mean parameters of each subsample in Table \ref{table_param} (not taking into account the doubtful afterglow of \object{GRB 020321}). We recall that the XMM-Newton and BeppoSAX samples are not very different in the observation time, while the Chandra sample is observed later. In both Tables \ref{table_flux} and \ref{table_param} this discrepancy between the XMM-Newton and Chandra samples appears in both the fluxes and the decay index. The higher flux extrapolated at 11 hours for the Chandra Imaging sample can be explained as due to the steeper decay of these bursts : the interpolation to early times is very sensitive to a significant change of the decay properties. We stress that the Chandra Grating data are not strongly affected by this effect due to the small change of the mean decay index compared to the mean XMM-Newton decay index (see Table \ref{table_param}). We note a marginal discrepancy not significant in the spectral index (the errors reported in Table \ref{table_param} are quoted to the $1\sigma$ level only) between the Chandra Imaging and the other samples. We also note that there is no strong difference between the XMM-Newton and BeppoSAX samples, which indicates that there is no selection bias between these two samples. Taking into account the flux level of \object{GRB 020321}, the mean XMM-Newton afterglow flux decrease to $10^{-12.7\pm 0.3}$ erg cm$^{-2}$ s$^{-1}$, still of the same order of magnitude of the BeppoSAX one.

\begin{table}
\caption{Mean parameters of each sub-sample from this paper. We include the decay index ($\delta$), the spectral index ($\alpha$), the flux at 11 hours after the burst and the observation start time. We also recall the main parameters from the BeppoSAX sample \citep{dep05}. All errors are quoted at a 68\% confidence level.}             
\label{table_param}      
\centering                          
\begin{tabular}{lcccc}        
\hline\hline                 
            & Chandra         & Chandra         &   XMM           & BeppoSAX  \\ 
            & Imaging         & Grating         &                 & \\
\hline                   
$\delta$    & $2.0 \pm 0.3$   & $1.4 \pm 0.1$   & $1.2 \pm 0.2$   & $1.3 \pm 0.1$ \\
$\alpha$    & $0.8 \pm 0.1$   & $1.0 \pm 0.1$   & $1.2 \pm 0.2$   & $1.2 \pm 0.1$  \\
Log(Flux)   & $-11.6 \pm 0.3$ & $-11.8 \pm 0.3$ & $-12.6 \pm 0.2$ & $-12.2 \pm 0.1$ \\
 @ 11 hours &                 &                 &                 & \\
Obs. time   & 145.4           &  95.9           &  30.1           & 36.5\\
(ksec)      &                 &                 &                 & \\
\hline
\end{tabular}
\end{table}

\begin{figure}
   \centering
   \resizebox{8.5cm}{!}{\includegraphics{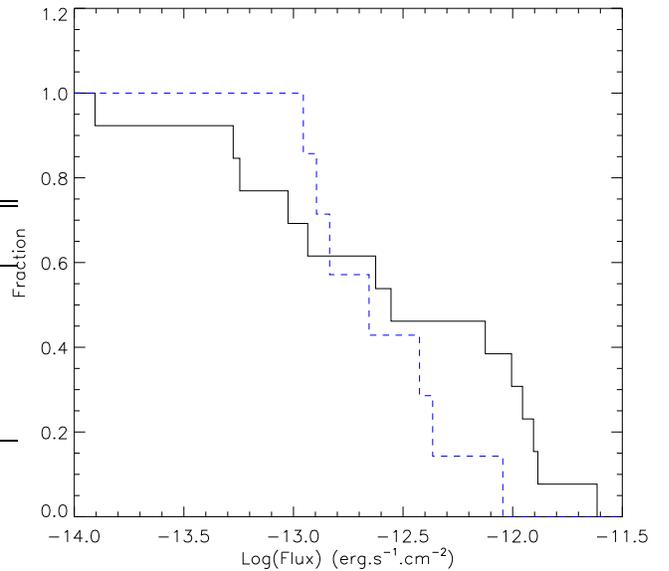}}
   \caption{Cumulative distribution of flux of the afterglows observed by XMM-Newton (dashed line) and Swift-XRT (continuous line) 12 hours after the burst.} 
              \label{fig_swift}%
    \end{figure}

\citet{ber05} found that SWIFT X-ray afterglows are fainter compared to those observed by other instruments. They explain this discrepancy by a selection effect on the instrument that detected the burst (BeppoSAX, IPN, INTEGRAL, HETE-2 versus SWIFT). They support this result using a dataset of 14 SWIFT bursts and 49 other bursts. To do so, they extrapolated (or interpolated) to 12 hours after the burst the flux quoted in published papers or in GCNs, using the reported decay index (when present) or a mean value of $\delta$ (1.3 if the quoted afterglow flux was measured at least one hour after the burst, or 1 otherwise), and a simple power law decay. We have checked if this discrepancy also appears using the limited sample of XMM-Newton and Chandra bursts (15 bursts).
In order to take into account the bias effect toward higher fluxes observed in the Chandra data discussed above, we removed the Chandra observations and retained only the 7 XMM-Newton bursts. We present in Fig. \ref{fig_swift} the comparison of these 14 SWIFT bursts with the 7 XMM-Newton ones. Four Swift bursts are observed with a flux larger than $10^{-12}$ erg s$^{-1}$ cm$^{-2}$, while no XMM-Newton bursts are observed at this level of flux. Six of the Swift bursts are brighter than $5 \times 10^{-13}$ erg s$^{-1}$ cm$^{-2}$ while only one XMM-Newton burst is observed at this level of brightness. It thus seems, using our very limited sample, that SWIFT bursts are not fainter than the XMM-Newton ones. Because the XMM-Newton sample is limited (7 bursts), we have checked if this result still holds using the more complete BeppoSAX sample. Comparing the mean afterglow flux measured by SWIFT, which is $10^{-12.4}$ erg cm$^{-2}$ s$^{-1}$ at 12 hours after the burst, with the BeppoSAX one ($10^{-12.2}$ erg cm$^{-2}$ s$^{-1}$, extrapolated to 12 hours using the main parameters listed in Table \ref{table_param}), we do not observe any strong discrepancy between the two samples.

 Moreover, we can consider the effect of the early afterglow evolution on the comparison. As reported by \citet{but05} and \citet{chi05}, some of the early afterglow light curves are not compatible with a simple power law. For this reason, we have discarded from the 14 SWIFT burst sample those for which the flux at 12 hours after the burst was extrapolated using an observation made at less than 3 hours after the burst. Four SWIFT bursts were removed. Those 4 bursts removed from the sample have the lowest extrapolated fluxes. We still do not observe significant discrepancies between SWIFT and BeppoSAX bursts : the mean SWIFT flux value changes to $10^{-12.3}$ erg cm$^{-2}$ s$^{-1}$ using only the reduced SWIFT sample.

\subsection{Time evolution of the closure relationships}
\label{sec_discu_jet}

   \begin{figure}
   \centering
   \resizebox{9.5cm}{!}{\includegraphics[bb=0 92 592 800]{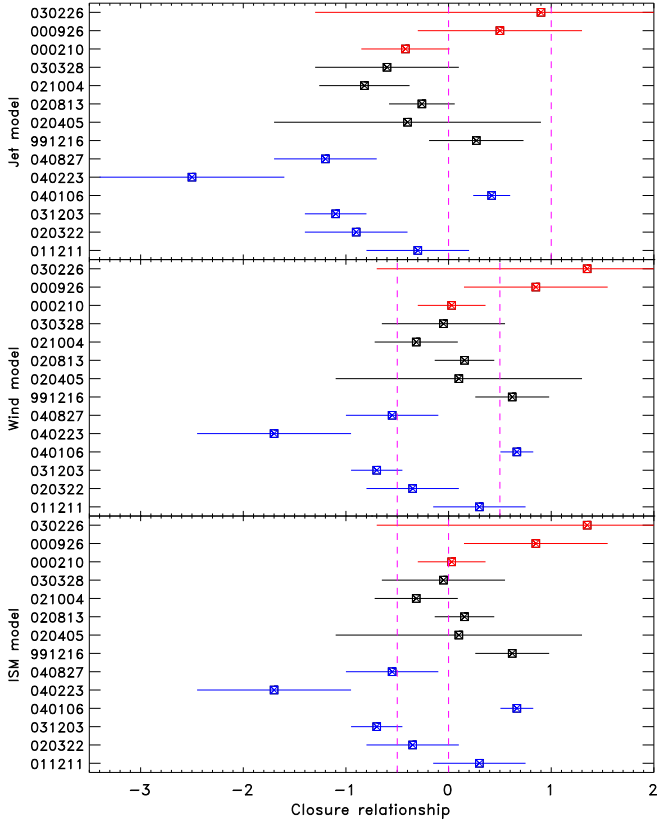}}
   \caption{The closure relationships for all bursts with constraints on both the spectral and temporal decay indexes. We indicate the closures for the three cases (Jet effects, Wind model, ISM model) in the three panels. Vertical lines indicate the theoretical expected values. Afterglows observed by Chandra are located at the top (in red for Chandra Imaging and in black for Chandra Grating data), while those observed by XMM-Newton are located at the bottom (in blue, see electronic version for colors).} 
              \label{Fig2}%
    \end{figure}

The values of the decay index ($\delta$) and the spectral index ($\alpha$) are linked through closure relationships. These relationships depend on the burst environment and the burst geometry \citep{rho97,sar98, che99}. According to \citet{sar98}, if the burst is surrounded by a medium with uniform density, then :

\begin{eqnarray}
\label{eq1}
\delta - 1.5 \alpha = -0.5 & & \nu > \nu_c \\
\label{eq1bis}
\delta - 1.5 \alpha = \phantom{-0.}0 & & \nu < \nu_c 
\end{eqnarray}

In the case of a surrounding medium with a non uniform profile decreasing like $r^{-2}$ \citep[the wind profile,][]{che99, che04}, we obtain :

\begin{eqnarray}
\label{eq2}
\delta - 1.5 \alpha = -0.5 & & \nu > \nu_c \\
\label{eq2bis}
\delta - 1.5 \alpha = \phantom{-}0.5 & & \nu < \nu_c 
\end{eqnarray}

One can also estimate the geometry of the burst by a set of closure relationships. In case of a jet, when the opening angle of the beam is smaller than or equal to the (time dependent) relativistic collimation, one obtains \citep{rho97} independently of the density profile :

\begin{eqnarray}
\delta - 2.0 \alpha = 0 & & \nu > \nu_c \label{eq3}\\
\delta - 2.0 \alpha = 1 & & \nu < \nu_c \label{eq3bis}
\end{eqnarray}

We present the closure relationships for all bursts with good constraints in Fig \ref{Fig2}. The average positions for each sub-sample are
$\delta - 2.0\alpha =$ $-0.93 \pm 0.36$, $-0.36 \pm 0.17$, $0.33 \pm 0.32$ and $-1.14 \pm 0.18$ for XMM-Newton, Chandra (grating), Chandra (imaging) and BeppoSAX \citep[see][for the BeppoSAX case]{dep05} respectively when equations \ref{eq3} and \ref{eq3bis} apply. When equations \ref{eq1}, \ref{eq1bis}, \ref{eq2} and \ref{eq2bis} apply, these positions are $\delta - 1.5\alpha =$ $-0.39 \pm 0.31$, $0.10 \pm 0.14$, $0.74 \pm 0.32$ and $-0.6 \pm 0.2$ respectively.

The XMM-Newton and BeppoSAX results are very similar. On the other hand, the mean position value changes from the XMM-Newton sample to the Chandra samples, with two extreme positions (XMM-Newton and Chandra imaging samples) and one intermediate position (Chandra grating). The mean observation start times within these three samples are 30.1, 95.9 and 145.4 kiloseconds for the XMM-Newton, Chandra grating and Chandra imaging samples. We can interpret this evolution either as a passing through of the cooling frequency in the X-ray band or as a jet signature. We discuss this in the following two sections, but note however that this is the first time we observe a clear evolution with time of the global afterglow properties from late X-ray data.

\begin{figure}
\centering
\includegraphics[width=9cm]{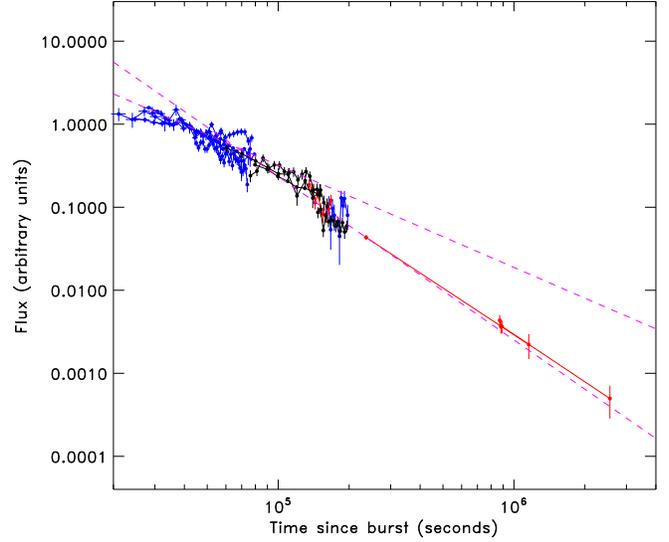}
\caption{Composite light curve of an X-ray afterglow. XMM-Newton bursts are in blue, Chandra Grating bursts in black, Chandra-Imaging bursts in red. Points relative to the same afterglow are connected by a solid line. The two dashed purple lines represent the mean decay observed from XMM-Newton and Chandra Imaging bursts.\label{figure_lc}}
\end{figure}

\subsection{Burst geometry}
\label{sec_geometry}

According to \citet{sar99} and assuming a burst located at z=1, the beaming angle is :

\begin{equation}
\label{equa_angle}
\theta = 0.166\left(\frac{n_0}{E_{i, 52}}\right)^{1/8} t^{3/8}_{b} \ \mathrm{rad}
\end{equation}

In Eqn. \ref{equa_angle}, $n_0$ is the density in particle cm$^{-3}$, $E_{i, 52}$ is the isotropic energy in units of $10^{52}$ erg, and $t_{b}$ is the jet break date expressed in days after the burst. The top panel of Fig. \ref{Fig2} displays the result for the jet signature, and indicates that we can rule out a jet signature within XMM-Newton bursts and cannot rule out this signature within Chandra bursts. This does not exclude the possibility of a collimated fireball in the XMM-Newton burst cases, but simply puts a lower limit on the jet opening angle. Other facts indicate a possible break in the Chandra burst light curves. As can be noted in Table \ref{table_flux}, the Chandra bursts appear brighter than the XMM-Newton ones when extrapolated back to half a day after the burst. Within the Chandra sample, there is also a trend : the brightest afterglows (12 hours after the bursts) are also the ones observed latest. All of this may indicate that the crude extrapolation using a single power law is not correct, and that there is a break in the light curve {\it before} the observation. Moreover, the position evolution with time of the calculated jet closure relationships (Eqn. \ref{eq3} and \ref{eq3bis}) as previously discussed can be interpreted as a convergence toward the expected value in the case of a jet signature. This is also clear if one produces a composite light curve from all the bursts observed (see Fig. \ref{figure_lc}). 

To produce Fig. \ref{figure_lc}, we first rescaled the flux levels of the XMM-Newton bursts observed before $10^5$ seconds to match the mean XMM-Newton flux level at 11 hours (which gives the value 1 in Fig. \ref{figure_lc}). We then scaled the Chandra grating flux levels to the mean Chandra grating flux level, and used a multiplicative factor to smoothly connect the XMM-Newton and Chandra grating light curves. We then rescaled the remaining bursts (one late XMM-Newton and 3 Chandra Imaging observations) to connect the composite light curve. One can clearly see an evolution from the XMM-Newton to the Chandra Imaging observations. The decay changes from $1.2 \pm 0.2$ to $2.0 \pm 0.3$ ($\Delta\delta = 0.8 \pm 0.5$, 90\% confidence level), consistent with a jet break (the theoretical value is $\Delta\delta = 0.83-1.33$, depending on the position of the cooling frequency and the burst environment) but not with a cooling break ($\Delta\delta = 0.25$). 
 We note however that some faint events feature a shallower apparent decay (e.g. \object{GRB 030723}), not compatible with a jet break. While we cannot exclude the hypothesis that some of these object with a shallow decay are not related to GRBs \citep[other galactic X-ray objects are variable, see e.g.][]{gen03} or indeed present a large beaming angle, this shallow decay may also indicate {\it flaring} afterglows like \object{GRB 970528} \citep{pir98} or \object{GRB 050904} \citep{wat05, cum05}. The light curve resolution does not allow us to discriminate between these 3 hypothese, and in the remaining we use only bursts with a good light curve resolution. Doing so, we
thus explain the evolution of the closure relationship values and the steep Chandra Imaging burst decays by a jet effect at late times. Assuming there is no selection bias within the XMM-Newton and Chandra Imaging samples, and taking into account the mean observation time (1 day after the burst and 2 to 10 days after the burst for XMM-Newton and Chandra respectively), we can set lower and an upper limits : 

\begin{equation}
\label{equa_angle2}
0.166\left(\frac{n_0}{E_{i, 52}}\right)^{1/8} < \theta < 0.39\left(\frac{n_0}{E_{i, 52}}\right)^{1/8}
\end{equation}

Assuming typical values for $n_0$ and $E_{i, 52}$ \citep[0.1, 10 respectively,][]{fra01}, we obtain from Eqn. \ref{equa_angle2} : $0.09$ rad $< \theta < 0.22$ rad ($5.1\deg < \theta < 12.6\deg$), which is in agreement with previous works \citep[e.g.][]{fra01, ber03}. 

The steep decay observed late can also explain the non-detections. The observations that led to non detections are performed very late (see Table \ref{table3}). At that time, one should observe a steep decay of the afterglow due to the jet effect. This could prevent a detection due to the afterglow faintness at that time.

\begin{figure}
   \centering
   \resizebox{9.5cm}{!}{\includegraphics[bb=10 92 592 600]{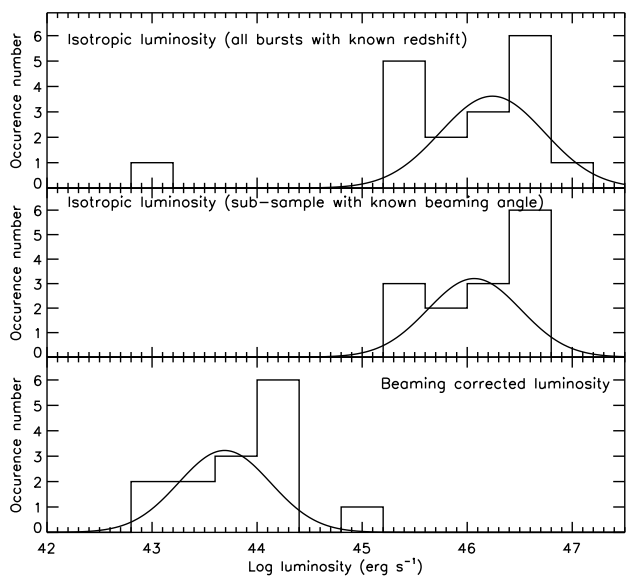}}
   \caption{Distributions of the isotropic luminosity for all BeppoSAX, XMM-Newton or Chandra bursts with good constraints on the spectral and temporal decay index and known redshift (top panel), for a sub-sample of bursts with known beaming angle (middle panel), and distribution of the beaming corrected luminosity of this sub-panel (bottom panel) using the method of \citet{ber03}. All luminosities were calculated at 11 hours after the burst (in the burst rest frame), using the observed values of spectral index, temporal decay, redshift and beaming angle. The low luminosity outlier in the top panel is \object{GRB 031203}, not plotted in the two other panels.} 
              \label{fig_berger}%
    \end{figure}

 We have looked for a standard energy emission in the X-ray afterglow as in \citet{ber03}.
As the XMM-Newton and Chandra catalog does not present enough bursts with known beaming angle to have meaningful results, we have extended our sample using the BeppoSAX catalog \citep{dep05}. We have computed the X-ray luminosity using the conversion formula quoted in \citet{lam00} and used by \citet{ber03}. We have included in our sample bursts neither with unknown redshift, nor bursts with poorly constrained spectral and temporal decay indexes. Eighteen bursts were included, out of which 14 have a known beaming angle. We have computed the luminosity at a common epoch of 11 hours after the burst using for each burst its spectral and temporal decay indices \citep[note that][used a mean spectral index and mean values for the temporal decay index and redshift for unknown values]{ber03}.
The result is shown in Fig. \ref{fig_berger}. In the top panel, we have plotted the isotropic luminosity distribution for all the 18 bursts indicated above. In the middle panel, we plotted the isotropic luminosity distribution for only those bursts with a known beaming angle (14 bursts), and in the bottom panel, we plotted the beaming corrected luminosity distribution of these 14 bursts.
We have fitted a Gaussian distribution to each luminosity distribution. We obtained $\sigma = 0.5 \pm 0.3$ for the upper panel, $\sigma = 0.5 \pm 0.2$ for the middle panel and $\sigma = 0.4 \pm 0.2$ for the lower panel.
The values obtained by \citet{ber03} were $\sigma = 0.7 \pm 0.2$ for the isotropic luminosity distribution and $\sigma = 0.3 \pm 0.1$ for the beaming-corrected luminosity distribution. 
We observe the same (within error bars) distributions when one corrects for beaming. However, our isotropic luminosity distribution is narrower than the one observed by \citet{ber03}, while compatible within the error bars. 
In fact, its width is consistent with that of the beaming-corrected distribution. So, we do not observe a significant shrinking of the luminosity distribution when taking into account the beaming.
We thus cannot confirm the \citet{ber03} claim for a standard release of energy in the X-ray afterglow.
The addition of several SWIFT bursts may reduce the error bars and may help to reach a conclusion on that effect.
 Also, we can observe in the top panel a bimodal distribution compatible with the results of \citet{gen05}.

\subsection{Surrounding medium}

The central and bottom panels of Fig. \ref{Fig2} display the results for the wind and constant density environments respectively. 

Most of the bursts can fit both of the medium classes. This is mainly due to the degenerations observed for 2 closure relationships, requiring data taken below $\nu_c$ to constrain the burst environment. However, we can distinguish the medium type for one burst.

Using the X-ray afterglow data only, we can constrain the burst environment of \object{GRB 040106} to be a wind environment \citep{gen04b}. \citet{mor05} ruled out this conclusion on the basis of their re-analysis, stating that the closure was at more than 3$\sigma$ of the theoretical position. They based their conclusions on the 1$\sigma$ error bars they quoted. Using our analysis and the 90\% confidence level as derived from a study of the $\chi^2$ space for all the errors (which is larger than twice the 1$\sigma$ error), we confirm that the theoretical position of the closure is within the 90\% confidence level region of the observed closure, and thus we claim that this burst is indeed surrounded by a wind profile due to the burst progenitor.

\subsection{Absorption}
\label{section_absorp}

\begin{figure}
\centering
\includegraphics[width=8cm]{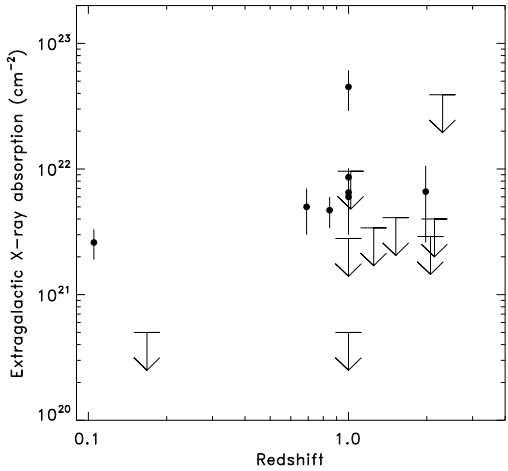}
\caption{\label{figure_absorb}X-ray absorption versus redshift. We indicate in this figure all detected absorptions or the corresponding upper limits for all bursts bright enough for a spectral study. The redshift was assumed to be one when not available.}
\end{figure}

From our analysis, 8 bursts (of 17) present an excess of absorption that ranges from $0.5 \times 10^{21}$ to $88 \times 10^{21}$ cm$^{-2}$, with a median value of $\sim 5.8 \times 10^{21}$ cm$^{-2}$. We present the distribution of absorption versus the redshift in Fig. \ref{figure_absorb}. \object{GRB 040223} is located near the galactic plane (l=341.6138$\degr$, b=3.1940$\degr$). The galactic column density in that direction is $6\times10^{21}$ cm$^{-2}$ \citep{dic92}, and varies (within one degree) between $4.46\times10^{21}$ and $9.26\times10^{21}$ cm$^{-2}$ (as one may expect, this increase is due to a change in the distance to the galactic disk). \citet{sch98} indicated that their work cannot be used with $|b| < 5 \degr$; as a diagnostic, the $N_H$ value expected from their map of dust is $\sim 9\times10^{21}$. Assuming that all the absorption is galactic, we obtain $N_H = (1.8 \pm 0.3)\times 10^{22}$, which is not compatible with either the \citet{dic92} or the \citet{sch98} values. Assuming the largest observed absorption value within one degree, one obtains an extragalactic absorption of $N_{H, z=1}= (4.5 \pm 1.5) \times 10^{22}$, which is larger than the values observed in other bursts. In the remaining, we will conservatively use this value. If this burst is at short distances ($z \sim 0.1$), then the extragalactic absorption value becomes of the same order as the one observed in other bursts.

We have investigated whether some non detections of the optical afterglow may be due to the absorption. Assuming a galactic gas to dust law, using the work of \citet{sch98}, the observed excess of absorption implies an extinction in the R band between 0.27 and 25.06 magnitudes (with a mean of 3.23 magnitudes, excluding the value of \object{GRB 040223} from the calculation due to its uncertainty). This high optical extinction is not in agreement with optical data and lead \citet{gal01} to propose that GRBs can destroy dust around them. On the other hand, \citet{str04} indicated that the galactic gas to dust law is not suited for GRB studies, and indicated that the correct laws may be those of \citet{cal94} and \citet{cal97}. Using these laws, the R extinction is still ranging from 0.13 to 11.08 magnitudes (with a mean of 1.43 magnitudes). A burst like \object{GRB 040223}, with an R extinction of at least 11.08 magnitudes, would not be detectable in the optical, even at very early times.

   \begin{figure}
   \centering
   \resizebox{7.6cm}{!}{\includegraphics{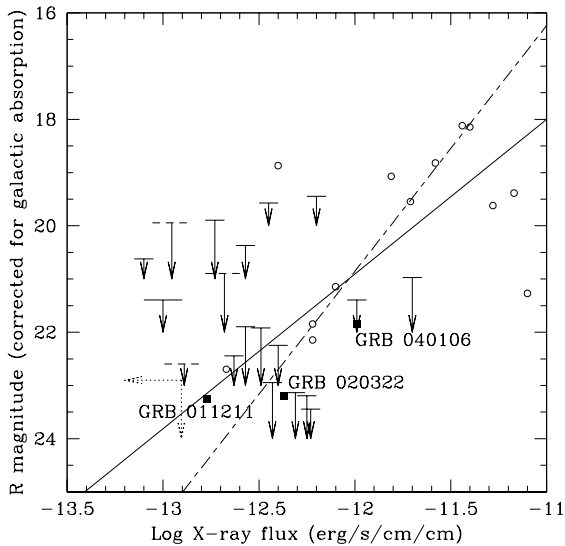}}
   \caption{Optical versus X-ray fluxes of GRB afterglows 11 hours after the burst. Squares and open circles represent XMM-Newton and Beppo-SAX data \citep[extracted from][]{pas03} respectively. Lines indicate the best fit relationships (see text for details).}
              \label{Fig1}%
    \end{figure}

Bursts with no optical detection but with a positive detection in X-rays or radio are called Dark Bursts. \citet{pas03} indicated that 75\% of dark bursts are faint bursts with a failed detection of the optical afterglow due to its faintness. The remaining ones are those that do not fit the simple fireball afterglow spectral model \citep[unabsorbed broken power laws, see][]{sar98} because of a depletion in the optical. According to \citet{pas03}, these {\it truly dark} bursts account for the remaining 25 \% of dark bursts. They can be distant bursts ($z>5$, so that the Ly$_\alpha$ break is redshifted into the optical range) or absorbed bursts (the effect of absorption would be more severe in optical than in X-rays or radio, and can prevent an optical detection). This result was confirmed by \citet{jak04} using a different method.

Eight bursts of our sample do not present an optical afterglow. \object{GRB 040223} is located near the galactic plane, and the galactic absorption can explain the non-detection, as indicated before. Two of the remaining 7 dark bursts are absorbed \citep[one of these two bursts has been classified as {\it truly dark} by][]{jak04}. This represents $\sim 28$ \% of the dark burst sample. We used a very limited sample, and that this result should be considered as a trend. This number is of the same order as the 25 \% of {\it truly dark} bursts (the counting statistic uncertainty is $\sim 14$ \%). The {\it truly dark} bursts may thus be only absorbed bursts, whereas we cannot exclude that a small fraction ($<\sim 14$\%) of dark bursts may in fact be distant (z$>5$) bursts.

We present the optical versus X-ray fluxes diagram in Fig. \ref{Fig1}. Because of the bias discussed earlier and the possible jet effects in the light curves, we used only the XMM-Newton sample, together with the BeppoSAX sample, to build this figure. All the values have been corrected for the galactic absorption (X-ray) or extinction (R band), using the work of \citet{sch98}. Not all XMM-Newton bursts are displayed in this figure : three GRBs in our sample do not have detected optical afterglows (or the detections are very late). We do not include the corresponding upper limits in Fig. \ref{Fig1} due to the poor constraints they give (they are R$<18$ at 11 hours after the burst once corrected for the galactic reddening). We have fitted these data points using an empirical power law. The best fitted relationships are indicated in the figure. To derive these relationships, we have included (dashed line) and excluded (solid line) from the fit the upper limits. We can use these laws to test the hypothesis of absorption around faint optical afterglows. To do so, we focus on \object{GRB 020322}. This burst is located in the ``dark'' side of the figure : it is a normally bright X-ray afterglow, while it displays a faint optical afterglow. It also displays an excess of X-ray absorption \citep[see Table \ref{table1} and][]{wat02b}. Using the laws of  \citet{cal94} and \citet{cal97}, we calculated an extinction of 1.6 $\pm$ 0.4 in the R band. The error on this value is calculated taking into account all the uncertainties of the spectral fit (i.e. on the galactic absorption, on the intrinsic absorption and on the spectral index). The best fitted relationships imply an extinction of $0.5-2.5$ for this burst, compatible with our finding from the X-ray. As other bursts present an extragalactic absorption higher than the one of \object{GRB 020322}, some of the dark bursts located in the same area of this diagram may be absorbed bursts. Note that this does not imply that all dark bursts are absorbed ones. In fact, $\sim 50$\% of bursts present an excess of absorption, and 5 of these are not dark : the presence of absorption does not conflict in all the cases with the optical afterglow detection.

\section{Conclusion}

We have presented a catalog of X-ray afterglows observed with XMM-Newton and Chandra. We have derived the absorption in the host galaxy, the decay and spectral indexes, and the observed unabsorbed fluxes. 

We have observed variations of the global properties of the X-ray afterglows within the samples that we explain by a jet effect in the sample. This is observed from both a variation of the closure relationships and the steepening of the mean decay index for bursts observed late. We have shown that jet effects are not present within one day after the burst. Taking advantage of the late observation time of Chandra, we find that the jet signature occurs between two and ten days after the burst. This allowed us to constrain the jet opening angle value. 
We have not noticed differences between the XMM-Newton and BeppoSAX samples that can be explained by an absence of a selection bias between these observatories. 
 We have not observed a significant shrinking of the luminosity distribution when we corrected for beaming. This result needs to be confirmed by more burst observations, e.g. by SWIFT.
In addition, we have found that the SWIFT X-ray afterglows are similar to those of XMM-Newton and BeppoSAX, while Chandra ones are biased toward higher luminosities. If this bias is not considered, the SWIFT bursts are fainter than the complete sample of all bursts listed here.
We have also indicated that several dark bursts may be simply extinguished in optical due to the high absorption observed around the bursts.

\begin{acknowledgements}
We thank G. Garmire for his help during the analysis of the Chandra grating data, and B. McBreen for pointing out some missed info in the catalog. We thank the anonymous referee for the report and for pointing out one missed burst. Thanks are also due to M. De Pasquale for help with the temporal analysis. This work was supported by the EU FP5 RTN 'Gamma ray bursts: an enigma and a tool'. A.C. acknowledges support from an INFN grant. This work is based on observations obtained with XMM-Newton, an ESA science mission with instruments and contributions directly funded by ESA Member States and NASA.
\end{acknowledgements}

\end{document}